
%
%



\magnification=\magstep1	
\raggedbottom

\parskip=9pt
\def\singlespace{\baselineskip=12pt}      
\def\sesquispace{\baselineskip=16pt}      

\overfullrule=0pt 
 



%
 \let\miguu=\footnote
 \def\footnote#1#2{{$\,$\parindent=9pt\baselineskip=13pt%
 \miguu{#1}{#2\vskip -7truept}}}
%
%


\def\=>{\Rightarrow}
\def\==>{\Longrightarrow}
 
 \def\dal{\displaystyle{{\hbox to 0pt{$\sqcup$\hss}}\sqcap}}
 
%
\def\lto{\mathop
        {\hbox{${\lower3.8pt\hbox{$<$}}\atop{\raise0.2pt\hbox{$\sim$}}$}}}
\def\gto{\mathop
        {\hbox{${\lower3.8pt\hbox{$>$}}\atop{\raise0.2pt\hbox{$\sim$}}$}}}
%
 



\def\to{\rightarrow}		

\def\tilde{\widetilde}		


\def\Reals{{\rm I\!\rm R}}	

\def				
  \Complexes
   {{\rm C}\llap{\vrule height6.3pt width1pt depth-.4pt\phantom t}}


\def\tr{\rm tr}			



\def\interior #1 {  \buildrel\circ\over  #1}     





\def\BulletItem #1 {\item{$\bullet$}{#1}}


\def\AbstractBegins
{
 \singlespace                                        
 \bigskip\leftskip=1.5truecm\rightskip=1.5truecm     
 \centerline{\bf Abstract}
 \smallskip
 \noindent	
 } 
\def\AbstractEnds{\bigskip\leftskip=0truecm\rightskip=0truecm}

\def\ReferencesBegin
{
\singlespace					   
\vskip 0.5truein
\centerline           {\bf References}
\par\nobreak
\medskip
\noindent
\parindent=2pt
\parskip=4pt			        
 }

\def\section #1    {\bigskip\noindent{\bf  #1 }\par\nobreak\smallskip}

\def\subsection #1 {\medskip\noindent{\it [ #1 ]}\par\nobreak\smallskip}

\def\eprint#1{$\langle$e-print archive: #1$\rangle$}

\def\A{{\cal A}}
\def\H{{\cal H}}
\def\P{{\cal P}}

\def\half{{\textstyle{1\over2}}} 

\def\tc{{t_c}}			
\def\sigc{\Sigma_c}		


\phantom{}
\vskip -1 true in
\medskip

\rightline{gr-qc/9902051}
\rightline{SU--GP--1998/12--1}

\vskip 0.3 true in

\vfill

\bigskip
\bigskip


\sesquispace
\centerline{\bf LARGE FLUCTUATIONS IN THE HORIZON AREA AND}

\centerline{\bf WHAT THEY CAN TELL US ABOUT ENTROPY AND QUANTUM GRAVITY}

\singlespace			        

\bigskip
\centerline {\it Rafael D. Sorkin}
\medskip
\smallskip
\centerline {\it Instituto de Ciencias Nucleares, 
                 UNAM, A. Postal 70-543,
                 D.F. 04510, Mexico}
\smallskip
\centerline { \it and }
\smallskip
\centerline {\it Department of Physics, 
                 Syracuse University, 
                 Syracuse, NY 13244-1130, U.S.A.}
\smallskip
\centerline {\it \qquad\qquad internet address: sorkin@physics.syr.edu}
 
\bigskip
\centerline {\it  Daniel Sudarsky}
\medskip
\centerline {\it Departamento de Gravitaci\'on y Teor\'\i a de Campos, 
  Instituto de Ciencias Nucleares} 
\centerline {\it Universidad Nacional Aut\'onoma de M\'exico,
	Apdo. Postal 70-543 M\'exico 04510 D.F, M\'exico} 
\smallskip
\centerline {\it \qquad\qquad 
internet address: sudarsky@xochitl.nuclecu.unam.mx}

\AbstractBegins 

We evoke situations where large fluctuations in the entropy are induced,
our main example being
a spacetime containing a potential black hole 
whose formation depends 
on the outcome of a quantum mechanical event.
We argue that 
the teleological character of the event horizon
implies that 
the consequent entropy fluctuations 
must be taken seriously 
in any interpretation of the quantal formalism.  
We then indicate how 
the entropy can be well defined 
despite the teleological character of the horizon, 
and we argue that 
this is possible 
only in the context of a spacetime or ``histories'' formulation 
of quantum gravity, 
as opposed to a canonical one,
concluding that 
only a spacetime formulation 
has the potential 
to compute ---
from first principles 
and in the general case ---
the entropy of a black hole.
From the entropy fluctuations in a related example, 
we also derive 
a condition governing 
the form 
taken by
the entropy,
when it is expressed 
as a function of
the quantal density-operator.
 
\AbstractEnds


 
\sesquispace

\section  {I. Introduction}

For a system in equilibrium, its thermodynamic entropy is by definition
a constant, but its statistical mechanical  ``entropy \`a la
Boltzmann'', being
a measure of the number of microstates making up the given ``macro'' or
``meso-''state, can, and in fact does fluctuate.  Since, in equilibrium,
the probability of a fluctuation associated with the entropy change
$\Delta S$ varies as $e^{- \Delta S}$, large fluctuations of
the entropy are very unlikely.  Nevertheless the existence, even of
small fluctuations  is important in principle, and in certain
situations they can give rise to readily observable effects like
critical opalescence.

Perhaps the most far reaching implication of entropy fluctuations on the
theoretical side is that the law of entropy increase can survive them as
an exact relationship at best in an average sense.  Given this, it seems
important that at least some averaged version of the entropy obey an
exact law of increase, because otherwise perpetual motion machines would
no longer be excludable on the basis of statistical mechanics.  But
whether or not the average entropy can decrease will depend on the
details of how the average is taken, and in this way we can obtain some
guidance as to which definition of the entropy is most appropriate.  For
reasons such as these, a situation where {\it large} entropy
fluctuations occur with high probability would seem to be of
particular interest.  In this paper, we will evoke such a situation.

We will in fact describe two such situations, one
involving a ``Schr{\"o}d\-ing\-er cat'' like apparatus, the other
involving a black hole and relying crucially on the teleological
character of its horizon --- the fact that it responds in a certain sense
to events that are still in the future.  Although the validity of the
first example can depend on how one interprets quantum mechanics, we
believe that that of the black hole example cannot.

The black hole example is important for another reason, 
as we will try to convince the reader.  
It illustrates the point that 
not every general relativistic body 
lends itself 
to description in terms of 
data on a spacelike hypersurface [R::forks].  
Rather, the teleological nature of
the horizon seems to demand 
description 
within 
a spacetime or ``histories'' framework, 
and indeed in a framework 
which extends those usually considered 
by incorporating histories 
that proceed into the distant future 
and then return to the present.

The detailed working out of these points will be found in the remainder
of the paper.  Section II poses the problem of how to define the entropy
in the face of its fluctuations; Section III analyzes the
``Schr{\"o}dinger cat'' example and shows how it can guide us in
selecting a suitable formula for the statistical mechanical entropy;
Section IV introduces our second example involving a quantum
mechanical event whose outcome will determine the formation or not of a
black hole, implying a macroscopically large fluctuation in the entropy
associated with hypersurfaces at slightly earlier times; Section V
schematically evaluates 
a certain expectation value of the area of the black
hole of this example within an extended path integral formalism; and
Section VI attempts (without full success) to justify the
identification of that expected area with the entropy of the black hole.
In Section VII, we discuss our results and offer some conclusions.

\section {II. The uses of large entropy fluctuations}

It is not our purpose here to delve deeply into the meaning of entropy
or to attempt to settle any of the persistent conceptual questions that
attach to that concept.
We merely want to recall some of the
relevant definitions in order to fix our terminology and help establish
a context for the ``Gedankenexperimenten'' we will consider.

Within classical statistical mechanics one can distinguish two, somewhat
different notions of entropy.  
The first, which we may call the ``Boltzmann entropy'', depends on the
exact microstate of the system under consideration and is defined as the
logarithm of the number of microstates belonging to the same equivalence
class as the given one.  
The equivalence relation here is that of being
``macroscopically indistinguishable'', 
and the equivalence classes may be called {\it mesostates}, 
using the terminology of [R::stoch-manifold] and [R::meso-def].  
If $N_i$ denotes the number of microstates making up 
the $i^{th}$ mesostate,\footnote{*}%
{Within classical physics, the counting of states or Boltzmannian
 ``complexions'' can be done, 
 at best, 
 up to an arbitrary normalization,
 of course.} 
then the Boltzmann entropy is,
$$
   S= \log N_i                             \eqno(E::SB)
$$
whenever the microstate finds itself within the $i^{th}$ mesostate.
Although one can argue\footnote{$^\dagger$}%
{e.g. using the fact that the time spent in the $i^{th}$ mesostate is
 proportional to $N_i$}  
that the Boltzmann entropy tends strongly to increase, any possibility
of formulating this tendency as an exact law is excluded by the
existence of downward fluctuations which normally are small, but which
also can be very large, albeit with what is usually taken to be
exponentially small probability.

In order to achieve an exact law of entropy increase despite the
fluctuations, one can attempt to replace the exact equation of motion
for the microstate by an approximate dynamics taking the form of a
Markov process on the space of mesostates (see [R::stoch-manifold] for
some references).  To the extent that such an approximation can be
justified one recovers an exact law of increase,
not for the Boltzmann entropy, 
but for what one may call
the ``Gibbs entropy'',
$$
    \sum_i p_i \log N_i + \sum_i p_i \log (1/p_i) ,  \eqno(E::Gibbs-S)
$$
where $p_i$ is the probability of realization of the $i$th state of the
Markov process (the $i$th {\it mesostate}).\footnote{*}%
{In writing the first term of (E::Gibbs-S), 
 we have assumed, 
 in accordance with Liouville's theorem, 
 that 
 the particular probability distribution given by
 $p_i\propto{}N_i$ 
 is time invariant 
 with respect to the Markov process.}
Notice that the Gibbs entropy is an ``ensemble functional'', rather than
a function of the actual physical microstate (or even mesostate) of the
system. 
Reinterpreted in terms of a probability density $\rho$ on the
space of {\it microstates}, it is (up to an additive constant)
$$
    \int dx \rho(x) \log (1/\rho(x)) ,          \eqno(E::SG-class)
$$
where the function $\rho$ is chosen to be constant within each mesostate;
i.e. $\rho$ is that function which correctly reproduces the overall
occupation probabilities $p_i$ of the mesostates and weights equally each
microstate within a given mesostate.  One sees from (E::Gibbs-S) that
the Gibbs entropy can be interpreted as the {\it average} value of the
Boltzmann entropy, 
$S_i=\log{N_i}$,
augmented by a term capturing the uncertainty concerning which mesostate
the system is actually in. (Conversely the Boltzmann entropy (E::SB) can
be interpreted as the Gibbs entropy evaluated on an ensemble distributed
evenly over the microstates of the $i^{th}$ mesostate.)

Thus, we obtain the second law of thermodynamics in an exact form, but
only at the double cost of (1) reinterpreting the increasing entropy as an
ensemble function rather that a state function, and (2) working with an
approximate dynamics rather than the exact one.
The passage from an exact micro-system to an approximating meso-system has
been called ``coarse graining'', and some such procedure seems always to
play a role in any derivation of the law of entropy increase, although
the ``meso-system'', need not always arise from so simple a
transformation as grouping the microstates into equivalence classes.
(Often it does arise this way, as when one defines an approximate
stochastic dynamics for a Brownian particle by averaging over the
degrees of freedom of the ``reservoir'' particles; here the position,
and possibly the velocity, of the Brownian particle serve to
parameterize the mesostate.
In contrast, there are other schemes, such as that of the ``BBGKY''
hierarchy, that don't clearly incorporate any well-defined space of
mesostates, or stochastic processes thereon, at all.  Particularly
interesting in this connection is the Boltzmann equation.  On one hand,
it might be viewed as a deterministic (and nonlinear!) dynamics on a
certain space of mesostates, the space of 1-particle distribution
functions.  On this view, the function $H$ of the so called
``$H$-theorem'' would be a species of Boltzmann entropy.  On the other
hand, the Boltzmann equation might alternatively be seen as first step
in the ``BBGKY hierarchy'', in which case it would be an equation of
motion for a probability distribution on a multi-particle phase space,
and the function $H$ would appear as a type of Gibbs entropy.)

For a quantum system the situation is similar, but with what would seem
to be a greater degree of conceptual confusion overall.
The actual counting of states is more satisfactory of course, but the
distinction between the Boltzmann and the Gibbs entropies seems less
clear cut, and depends partly on how one interprets the quantum
formalism.
There is disagreement, for example, over whether the classical contrast
between microstate and ensemble does or does not correspond to the
quantum distinction between state vector $\psi$ and density operator $\rho$.
Also the meaning of the coarse graining leading to the concept
of mesostate seems more obscure.  For example, if one chooses to
identify the state-vector $\psi$ with a microstate and some family of
subspaces of the quantal Hilbert space with the space of mesostates,
then one must deal with the fact that $\psi$ will almost never be ``in''
any single mesostate.  In the following , we will ignore such
subtleties as much as possible, and content ourselves with the fact
that fairly naive definitions of the Boltzmann and Gibbs entropies will
suffice for most of the questions we deal with. (We also will not
always specify Gibbs vs. Boltzmann entropy when the context does not
require it.)

The most obvious quantum analog of the integral expression
(E::SG-class) is the familiar operator expression,
$$
    S^{(1)} \, = \, \tr(- \rho \log \rho )   \,,    \eqno(E::S1)
$$
$\rho$ being the density operator. In thermal equilibrium, as
represented by the ``canonical ensemble'', $\rho=Z(T)^{-1}\exp(-H/T)$, 
(E::S1)
yields an entropy for which the First Law of Thermodynamics is
recovered as the exact relation
$$
       d<H> = TdS^{(1)} \,,
$$
where $<H>=\tr(\rho H)$.  The definition (E::S1) therefore gives a
consistent account of the entropy in equilibrium.  The principal
importance of entropy, however, stems not from its behavior in
equilibrium but from the way that it governs the {\it approach to
equilibrium} in accordance with the second law of thermodynamics.  That
is, its importance lies in its tendency to increase with time.

Almost as familiar as the expression (E::S1), 
however, 
is the fact that
its increase is incompatible with unitary evolution.  
Much as in the classical setting recalled above, 
the usual response to this difficulty is 
to attempt to modify 
either 
the expression for $S$ 
or 
the law of evolution of $\rho$, 
or both.  
One introduces some type of ``coarse graining'' 
into the description of the system, 
leading to 
a new $\rho$ acting in a new Hilbert space 
and to 
an approximate equation of motion 
for this coarse-grained $\rho$ 
which is nonunitary.
[ The one coarse graining known to us 
 where the new equation of motion is not obviously approximate 
 is that consisting in
 ignoring everything veiled by the surface of a black hole (event horizon).  
 Such a coarse graining is also
 less ``subjective'' than most others.
 These features are 
 part of 
 what makes the black hole case so interesting. ]
As long as the new evolution remains linear, there are general theorems
guaranteeing the increase of $S(\rho)$, under the condition that the
``totally random'' $\rho\propto 1$ is a fixed point of the evolution
[R::ref-on-increase]
[R::Raf2].  
Thus one recovers an exact, quantal second law
for the coarse-grained entropy $S^{(1)}$.  The fact that we obtain an
exact law suggests that (E::S1) in this case should be interpreted as a
Gibbs entropy (with $\rho$ interpreted as describing an ensemble.)
Recall however that 
(E::S1) 
will reduce to a Boltzmann entropy if we choose
an appropriately ``reduced" or `` collapsed" operator $\rho$.

Now the definition (E::S1) might look  both familiar and natural, however,
if all we demand of our definition of entropy is that its value
increase with time, then the same theorems that guarantee the increase
of (E::S1) work also for many alternative
expressions, including all of the form
$$
  S^{(2)} =   \tr \, F(\rho)           \eqno(E::S2)
$$
where $F(x)$ is an arbitrary concave function [R::Raf2].\footnote{$^\dagger$}
{The following theorem is proved implicitly in [R::Raf2], although 
the statement of the theorem there is limited to the case of
$F(x)=-x\log{x}$.
Let $\H$ be a finite dimensional Hilbert space and 
let $\P$ be the space of positive hermitian operators 
(density matrices)
on $\H$.  
Let $T$ be a linear, 
trace-preserving 
map of $\P$ into itself which
has the identity operator $1$ as a fixed point.  
Then, for any convex function $F:[0,1]\to\Reals$ 
and any (normalized) density matrix $\rho$, 
we have
$\tr{F(\rho)}\le\tr{F(T(\rho))}$.   
This implies 
the law of increase 
of the entropy (E::S2) 
(at least in finite dimensions)
if the following conditions hold: 
(i) evolution $\rho(t)\rightarrow\rho(t + \Delta t)$ 
is a linear, trace preserving map on the space of density matrices;
(ii) The ``maximally random state'', $\rho = 1 / (\tr{1})$ 
is preserved by the time evolution;
(iii) the evolution is Markovian 
in the sense that
$\rho(t+\Delta t)$ can be determined from $\rho(t)$ 
at any single earlier time $t$.}
Similarly the expression
$$
    S^{(3)} =   - \log \tr \rho^2  \eqno(E::S3)
$$
has also been suggested as an alternative and can be shown to be
nondecreasing under very general assumptions.  This last alternative is
particularly attractive since it is additive for uncorrelated subsystems
and takes the value $\log N$ for a macrostate composed of $N$
equiprobable microstates.  How, then, is one to select the correct
entropy?

In practice, the issue is completely irrelevant in most cases because,
in some imprecise sense, a single mesostate contains almost all the
populated microstates, fluctuations away from that mesostate are
negligible, and all entropy expressions 
(of both the Gibbs and Boltzmann variety) 
give
equivalent answers.  In principle, however, 
there can exist
situations where arbitrarily large downward fluctuations in the
(Boltzmann) entropy are highly probable, and these situations can provide
guidance in the selection of the correct expression for the Gibbs
entropy.

\def\tc{t_c}

The type of situation we have in mind involves very large fluctuations
arising from the amplification of suitable quantal events.  
Consider, for example, a flammable substance inside a box, completely
isolated from the outside world, with a quantum mechanical device that
acts as a trigger to ignite the substance.  The trigger might contain,
for example, a half-silvered mirror and a device which will project a
single photon at the mirror at some definite time $t = t_0$, in such a
way that the photon will do nothing if reflected, but will cause the
substance to be ignited if transmitted.

Let the entropy  (E::S1) of the contents of the box before $t_0$ be $S_0$, 
and let the resulting entropy corresponding to the case where the
substance does not (respectively does) ignite
be $S_I$ (respectively $S_{II}$).  
Clearly $S_0\approx S_I\ll S_{II}$.
Then, for an external observer,
the entropy at time $\tc>t_0$, but still before the box
is opened, must, according to (E::S1),
be taken to be $S(\tc)= \half S_I + \half S_{II} + \log 2$, 
where we have assumed that the probability of reflection at the
half-silvered mirror is $\half$.

Now, let's consider what happens when the box is opened at $t_2>\tc$.
With probability $1/2$,  the 
substance will have burned, whence
we will associate with the box an entropy 
$S_{II}$, and we will be in the ordinary situation where
the (Boltzmann) entropy has 
increased: 
$S(t_2)-S(\tc)\approx \half(S_{II}-S_I)\gg 1$.  
However, also with probability
$1/2$,  the 
substance will not have burned, whence we
will have to assign the box an entropy $S(t_2)=S_I$.
So this time, 
$S(t_2)-S(\tc)= \half(S_1-S_{II})\ll -1$, and the entropy will have
decreased substantially!  This
seemingly paradoxical 
behavior 
is reconciled with our expectations when
we notice that on average the entropy doesn't change\footnote{*}
{In saying that it doesn't change, we neglect the 
 small decrease by $\log{2}$ 
 associated with the fact that the twofold alternative of
 burned vs. not burned has been resolved by opening the box. Properly
 speaking we should, in order to keep to a consistently defined Gibbs
 entropy, continue to retain both the burned and unburned cases in the
 ensemble 
 (with, of course, the observer incorporated into the system), in which
 case the missing $\log{2}$ would be restored.  (We are dealing here
 with essentially the same reduction in entropy that ``Maxwell's demon''
 produces, and any analysis of that situation in relation to the second
 law should apply here as well, cf. [R::OPen].)}:
$<S(t_2)> = \half S_I + \half S_{II} \approx S (\tc)$.
In sum, the law of entropy increase holds on average,
but there can be arbitrarily large fluctuations in individual cases.

We are used to seemingly paradoxical situations arising when
Schr\"odinger's cat puts in an appearance, and so one might worry
that a better interpretation of the quantum formalism would vitiate the
account given above.
For example, 
one might think that the entropy at time $\tc$ is really zero 
because the system is still in a pure state, 
or, 
conversely,
one might think that interactions with the environment
actually ``reduce the wave function'' at time $t_0$, 
whence the entropy at $\tc$ would already be the same as it is at $t_2$.
Or one might worry that the entropy that fluctuates 
is merely some subjective entropy ``for us'', 
the external observer, 
rather than an entropy associated objectively with the system itself.
These worries disappear, 
we claim, 
and the situation therefore becomes more dramatic, 
when it involves black holes, 
thanks to the teleological nature of the event horizon.

\def\tc{t_c}

We refer here to the fact that the size and location of the event
horizon, on a given hypersurface $\Sigma$ is determined by events
occurring to the future of $\Sigma$.  Consider, for example, a
situation where, at some moment of time $\tc$ 
corresponding to a hypersurface $\sigc$, 
a Schwarzschild black hole of mass $M$ is present,
perturbed by the presence outside the horizon of a static body of mass
$m$, sustained, by a cable; and an individual is holding an ax with
which he or she might decide at $t = t_0 > \tc$ to cut the cable.  The
area of the event horizon on $\sigc$ will be $A_1$ (which can be
well approximated by $16\pi{G^2}{M^2}$) if our ax wielder decides not to cut
the cable, but it will be a larger area $A_2$, in between $A_1$ and
$16\pi{G^2}(M + m)^2$, if s/he does decide to cut it.  The main point is that
{\it this value can not be ascertained without knowledge of events
occurring to the future of} $\sigc$.  The ax handler, whose
decisions we might feel uncomfortable describing physically, can of
course be replaced by a quantum mechanical\footnote{$^\dagger$}
{It is not obvious to us that the random choice must be made on the quantum
level, but it does seem safer to use a quantum event, since, on current
understanding, the resulting choice is absolutely unpredictable by
anything existing on or to the past of $\sigc$.}
device, 
as in the previous example, 
which will induce 
a measurement-like event 
at $t=t_0$.  

Just as in the previous example, 
we can expect to obtain 
large fluctuations 
of either sign 
in the entropy.  
Indeed it is natural 
to assume 
that the entropy on $\sigc$ 
is given by
$$
   S(\tc) = q (2\pi A_1/\kappa) + p (2\pi A_2/\kappa)   \eqno(E::SBH1)
$$
where $\kappa=8\pi G$ and where $p$ is the probability that the quantum
mechanical ``choice'' 
to be made
at $t_0$ 
will result in the cable being cut,
and $q = 1-p$ 
is the probability of it not being cut.  
Below, we will attempt to demonstrate
that the entropy (E::S1) is in fact well-defined in this situation and
that its value is correctly given by eq. (E::SBH1).  
The large fluctuations then 
follow as before.\footnote{*}%
{Macroscopically large fluctuations can also occur in conjunction with
 phase transitions.  There, however, the fluctuations connect states of
 {\it equal} (total) entropy.  The thermal $\rho$ in such a case is
 spread out over many mesostates, illustrating clearly the sense in
 which $\rho$ may be said to describe an ``ensemble'' rather than an
 individual system, and in which, consequently, the entropy associated
 to such a $\rho$ is a Gibbs rather than a Boltzmann entropy.}

Finally, we will argue that this type of situation, resulting from the
teleological character of the event horizon, will be extremely
difficult, if not impossible, to describe and analyze in terms of state
vectors associated with hypersurfaces, as in canonical formalisms for
quantum gravity, and that the only framework that seems capable of
treating them 
is a path integral or ``history'' one.
In addition,  we remark here that the
black hole example illustrates, once again,
the untenability of a 
semiclassical version of gravity
like a theory based on $G_{ab}=\kappa<T_{ab}>$,
since  that would put the  horizon in the wrong place, no matter what
happens.

\section{III. Further analysis of the box example:
              Which entropy does Schr\"odinger's cat like best?}

\def\tc{t_c}

Consider 
once again 
our box $B$ containing an incendiary device $E$ connected to a
quantum mechanical trigger $Q$ 
(say a partially silvered mirror with transmission probability $p$, 
or a spin-$1/2$ particle with its spin in the $x$ direction 
which at $t=t_0$ is going to be subjected 
to a measurement of it's spin along the $\theta$ direction 
such that the result ``$+$'' will occur with probability $p$ 
and the result ``$-$'' will occur with probability $q=1-p$) 
in such a way that 
with probability $q$ the device will be ignited at time $t=t_0$, 
and with probability $p$ it will not be ignited.

To simplify matters we will suppose that initially the box's contents
$E+Q$ can be represented by a single quantum state-vector $|b>$ and
that, after the burning (if it occurs), the contents can be represented
(in virtue of suitable coarse graining, possibly involving
ultra-weak environmental influences) by a single mesostate comprising
a large number of equally
probable quantum state-vectors $|a_i>$, $ i= 1.....N$, representing the
possible microstates of the resulting hot gas and ashes.

Now let's suppose as in the original Schr{\"o}dinger's cat
Gedankenexperiment, that we do not open the box until a much later
moment $t=t_2$.  At any time $ \tc\in(t_0,t_2)$ the state of
the box's contents will be represented by the density matrix,
$$
 \rho(\tc) = p |b><b| +
 (q/N) \lbrace |a_1><a_1| +|a_2><a_2|+....+|a_N><a_N| \rbrace
 \eqno(E::state-t1)
$$
At any time $t$ 
after the box is opened, 
$t\ge{}t_2$,  
we will have two possible situations:
With probability $p$, 
the device will have remained unignited 
so the contents of the box will be described by 
$$
   \rho(t_2)^{(I)} =  |b><b| 
   \eqno(E::state-t2I)
$$
With probability $q$, the device will have ignited,
so the contents of the box will be described by
$$
\rho(t_2)^{(II)} =(1/N) \lbrace |a_1><a_1| +|a_2><a_2|+....+|a_N><a_N| \rbrace
\eqno(E::state-t2II)
$$

Now let's see what has happened to the entropy in each case.
To start with, the entropy at $t=\tc$ is
$$
       S(\tc) = - p \log (p) -q \log (q) + q \log (N) \approx q \log N ,
   \eqno(E::St1)
$$
while at $t=t_2 $ we have, with probability $p$,
$$
     S(t_2)^{(I)} = 0
$$
and with probability $q$,
$$
    S(t_2)^{(II)} = \log (N)
$$
Thus the entropy can either increase by $p\log N$
or decrease by $q\log N$
in the process of opening the box.
On the other hand the average  entropy is just
$$
   <S(t_2)> = q \log (N)
$$
which differs from $S(\tc)$ only by the quantity $S'=-p\log(p)-q\log(q)$,
corresponding to the fact that the alternative that is  
unresolved at $\tc$ is resolved at $t_2$ with a corresponding gain of
information.

Two aspects  of the above example are worth pointing out.
First, the fluctuation of the entropy is large:
$$
   \Delta S(t_2) = (<S(t_2)^2>-<S(t_2)>^2)^{1/2} = \sqrt{pq} \ \log (N)
$$
i.e. is of the order of $<S(t_2)>$ itself.  And second, the requirement
that the 
change in the average entropy after the box is opened arise only from
the information gained about the resolution of the initially unresolved
alternative is enough to single out almost uniquely the expression
(E::S1) for the entropy.

To see this let's compute the entropy of the system at $\tc$ using,
instead of (E::S1), the generic formula $S^{(2)}$ of (E::S2).  The
result is
$$
   \tilde S(\tc) = F(p) + N F(q/N)
$$
At $t=t_2$ we have, using again (E::S2), that with probability $p$,
$$
   \tilde S(t_2) = F(1)
$$
and with probability $q$,
$$
  \tilde S(t_2) = N F(1/N)
$$
Thus the average  entropy is just
$$
    <\tilde S(t_2)> =  p F(1) + q N F(1/N)
$$
Hence the change of entropy associated
with the opening of the box is 
$$
  \delta<\tilde S> =  p F(1) + q N F(1/N) - F(p) - N F(q/N)  \eqno(E::dTS)
$$
We expect that, since
the only change occurring with the opening of the box is the resolution
of the alternative related to the action of the quantum 
trigger 
$Q$, the
change in entropy can depend on $p$ and $q$ but not on the specific
nature of the incendiary device, so $\delta<\tilde S>$ must be
independent of $N$.  
Requiring this, 
we obtain
$$
  {{\partial \delta<\tilde S>} \over {\partial N}} =
   q F(1/N) -F(q/N) -(q/N) [ F'(1/N) - F' (q/N)]=0
   \eqno(E::eqF)
$$
where the prime denotes differentiation with respect to the function's
argument.  Putting $x\equiv 1/N$, taking the derivative of (E::eqF) with
respect to $q$, and setting $q=1$, we obtain
$$
   x^2 F''(x) -x F'(x) + F(x) =0 \eqno(E::eqF2)
$$
The general solution of this differential 
equation is $ F(x)= C_1 x \log (x) + C_2 x $
 where $C_1, C_2$ are arbitrary constants.
 Therefore the expression for the entropy becomes:
$$
  \tilde S = \tr(F(\rho)) = C_1\ \tr (\rho \log (\rho)) + C_2
\eqno(E::NewS)
$$
where we have used the fact that the density matrix is normalized to
$\tr(\rho)=1$.  It is also easy to see that the expression  $S^{(3)}$ of
(E::S3)
does not satisfy our condition of $N$-independence; it therefore is also
ruled out as an alternative to (E::S1).  Thus, the analysis of this
Gedankenexperiment has led us, on the basis of very natural
requirements to a unique expression (up to normalization and an additive
constant) for the entropy of a system described by the density operator
$\rho$. 
 
\section {IV. The example involving gravitational collapse}

\def\tc{t_c}
\def\Tc{T_c}

Consider a static spherically symmetric thin shell of mass $M$, in a
similarly static, spherically symmetric asymptotically flat spacetime.
The shell is fitted with a quantal device that at time $t=0$ 
(according to an internal clock) 
will make a random choice
between triggering or not the collapse of the shell, which would result
in the formation of a black hole.  More concretely, imagine the shell as
made of two thin massless concentric spherical reflecting walls
separated by a small distance, with electromagnetic radiation confined
between them.  The triggering device may be imagined as before, except
that this time, instead of igniting a fire, it makes
the internal wall transparent to radiation when it is activated.
Thus, with probability $1/2$ (say), our static shell of radiation
will become a null collapsing shell at $t=0$, resulting 
in the subsequent formation of a black hole with
mass $M$.  

One could worry about the feasibility of synchronizing the change in the
transparency of the different parts of the internal spherical wall
without having to delay the collapse for a time comparable to the light
travel time across the shell after the quantum mechanical decision has
been made.  We do not believe this poses a problem. We can easily
synchronize the change in, say, two opposite parts of the shell by using
an EPRB device as our trigger mechanism: Take a spinless particle at the
center of the shell which decays into two photons.  Let's fit the
internal wall with detectors that will measure the helicity of the
photons and give each of them instructions to change the transparency if
the photon it detects has positive helicity, but not if it has negative
helicity.  In this way, a coordinated collapse of the shell will start
at opposite points in the shell, without the need to propagate signals
across the shell after the quantum choice is made.  One can readily
extend this synchronization from a pair of antipodal points to the
entire shell by means of correlated many-particle states.

The metric outside the thin shell is, of course, the Schwarzschild
metric:
$$
     ds^2= -(1-2GM/r) dt^2 + (1-2GM/r)^{-1} dr^2 + r^2 d\Omega^2
     \eqno(E::metric1)
$$
for $r\geq R_{shell}$, and the metric inside it is the Minkowski metric:
$$
    ds^2= - dT^2 + dR^2+ R^2 d\Omega^2            \eqno(E::metric2)
$$
for $R\leq R_{shell}$.  
We approximate the shell as infinitely thin.
The matching of the exterior coordinates $(t,r)$ with the interior
coordinates $(T,R)$ can be deduced from the requirement that the metric
induced on the shell from the exterior spacetime must coincide with that
induced from the interior spacetime; while the trajectory of the shell
(in the case where it does move) can be deduced from the requirement
that it move at the speed of of light, or more formally, from the
requirement that the induced metric thereon be degenerate [R::Isr].

Let the motion of the shell be given by specifying the functions
$r_{shell}=R^{(1)}(t)$, 
in terms of the exterior coordinates, $r$, $t$, 
and
$R_{shell}=R^{(2)}(T)$, 
in terms of the interior coordinates, $R$, $T$. 
For the metric induced from the exterior spacetime, we have
$$
  d\sigma^2 
  = 
  - [(1-2M/r)-(1-2M/r)^{-1}(dr/dt)^2] dt^2 + r^2 d\Omega^2  \,,
  \eqno(E::imetric1)
$$
while 
for the metric induced from the interior spacetime, we have
$$
   d\sigma^2 = - [1 - (dR/dT)^2] dT^2+ R^2 d\Omega^2  \,,  \eqno(E::imetric2)
$$
(where we have taken $8\pi G=8\pi$). 
Agreement of these expressions requires 
$r_{shell}=R_{shell}$, 
or in other words, 
$R^{(1)}(t)=R^{(2)}(T)$,
together with
$$
  [(1-2M/R)-(1-2M/R)^{-1}(dR/dt)^2] dt^2 = [1 - (dR/dT)^2] dT^2   \eqno(E::Tt1)
$$
These matching conditions,
which let us relate the interior to the exterior coordinates,
work out slightly differently in the timelike
and null cases.

Let's choose our coordinates so that the ``moment of decision'' 
is $T=t=0$.  
Then for $t,T<0$ the shell is static 
and we have $R=R_0$, 
where $R_0$ is the initial radius of the shell.
In this case, we obtain from (E::Tt1)
$$
     T= \sqrt{1-2M/R_0} \ t   \eqno(E::Tt2)
$$

If the shell fails to collapse, then (E::Tt2) remains true for all time.
On the other hand, if the shell collapses as a null shell starting at
$t=0$, we can obtain (for $T$,$t\geq 0$), both $T$ and $t$ as functions
of $R$ from the condition that both sides of (E::Tt1) vanish, i.e. that
the induced metric on the shell be degenerate. From the right hand side
of this equation we obtain $R(T) = R_0-T$, and from its left hand side
we find
$$
     R(t) - R_0 + 2M \log( {R(t) - 2M \over R_0 - 2M} ) = -t
$$
(where we have used the initial condition $R(t=0) =R_0$).

Obviously, the collapsing shell will cross  the Schwarzschild radius
at $t=+\infty$, $T=R_0-2M$.
But, in fact, 
the horizon  will be formed earlier than that.  
To determine when,
consider a light signal starting at the center of the shell at $T=T_1$
and traveling radially outwards.  It will be 
able to escape to infinity iff it reaches the shell before the
collapse has occurred.  That 
is, it must reach the
shell while one still has $R_{shell}>2M$. 
The signal travels according to
$R=T-T_1$, whence it will meet the shell 
when $T-T_1=R(T)=R_0-T$,  that is to say, at
$T=(1/2)(R_0 +T_1)$, 
at which time
$R_{shell}=(1/2)(R_0-T_1)$.  
So, the signal will escape 
iff  $T_1<R_0-4M$.  
If we take, for example, 
the initial shell radius 
to be $R_0=3M$, 
then the signal must leave the center with $T_1<-M$ to be able
to escape, and the origin at $T >-M $ is already inside the horizon,
if it turns out that the collapse is in fact triggered at $T=0$.

To summarize, the locus of the horizon at times earlier than $T=0$
depends on what happens at $T=0$.  If there is no collapse, there is of
course no horizon.  If the collapse occurs at $T=0$ then the locus of
the horizon at earlier times is given by 
$$
     T - R = R_0 - 4M.
$$

Consider now 
a Cauchy hypersurface $\Sigma_{\tc}$ 
with $\tc<0$ defined 
by the condition $t=\tc$ 
outside the shell 
and by the
corresponding condition $T=\Tc=(1-2M/R_0)^{1/2}\tc$ 
inside the shell. 
What is the area $A_{\tc}$ 
of the intersection of the horizon with $\Sigma_{\tc}$? 
If we choose $\tc>-(4M-R_0)(1-2M/R_0)^{-1/2}$, 
and at $t=0$ 
the collapse is in fact triggered, 
we will have 
$$
  A_{\tc} = 4\pi R_c^2 = 4\pi (4M - R_0 + (1-2M/R_0)^{1/2}\tc)^2 
  \eqno(E::Atc1)
$$
For example, 
if we choose $R_0=3M$, 
we will have a nonvanishing area
for $\tc>-\sqrt{3} M$ 
(if the collapse is triggered)  
and its value will be
$$
    A_{\tc} =  4\pi (M + \tc /\sqrt{3})^2  \ ,           \eqno(E::Atc2)
$$
so 
for $\tc=-{\sqrt{3} \over 2} M$, 
say,
we will have $A_{\tc}=\pi M^2$.  
Of course,
if at $t=0$ 
the collapse is not triggered, 
we will have $ A_{\tc}= 0$.
Note that by taking $R_0$ 
sufficiently close to $2M$ 
we can have\footnote{$^\dagger$}%
{Assuming that it is possible, in principle, to build a shell
 arbitrarily close to the Schwarz\-schild radius.}
a nonzero intersection of $\Sigma_{\tc}$ 
with the horizon 
as early as desired in exterior time $\tc$; 
however, 
we will always have $\Tc > -(4M-R_0) >-2M$ 
when we have such a nonzero intersection.  
In all these situations, 
the area $A_{\tc}$ 
will be bounded by $16\pi{M^2}$, of course.

We have considered as a natural choice, the foliation of the region of
spacetime prior to $t=0, T=0$ by hypersurfaces $\Sigma_{\tc}$ that are
orthogonal to the timelike Killing field present in this region
($({{\partial}\over{\partial t}})^a$ outside the shell, and
$({{\partial}\over{\partial T}})^a$ inside).  Other, equally natural
foliations exhibit the same anticipatory behavior of the horizon.
Consider, in particular, the foliation by 
hypersurfaces $\Sigma^*_{\tc}$ which coincide with the previous ones
outside the shell, but are continued inside the shell as unions of
radially ingoing null geodesics.\footnote{*}%
{An ingoing null surface is particularly apropos in connection with the
 scheme proposed in [R::Raf2] and [R::chandra-adelaide]
 for actually
 proving, from first principles, the non-decreasing character of an
 entropy like that discussed in section VI below.
 We presume that, in fact, the quantum gravitational second law will
 require the entropy to increase along an arbitrary well-defined
 foliation, just as the classical second law does.}
Such a $\Sigma^*_{\tc}$ will be
defined in the outside by the condition $t=\tc$, and in the inside by
the condition $T+R=\Tc+R_0=(1-2M/R_0)^{1/2}\tc +R_0$.  If the collapse is
triggered at $t=0$, then the intersection of $\Sigma^*_{\tc}$ with the
horizon will have a radius
$$
    R_c^*= (1/2)((1-2M/R_0)^{1/2}\tc + 4M) \ , \eqno(E::Rc)
$$
as long as $\tc > -4M (1-2M/R_0)^{-1/2}$, 
and an area $A^*_{\tc} = \pi [(1-2M/R_0)^{1/2}\tc + 4M]^2$.  

In concluding this section we would like to emphasize the apparently
objective character of the ambiguity in the magnitude of the horizon
area in the above example.  It is not that ``we avoid finding out''
whether a horizon exists (as with the Schr{\"o}dinger cat example), but
that it is objectively impossible for anyone to find out, given access
only to information available on the given hypersurface.

 \section  {V. Path integral evaluation of the expected  horizon area}

In this section we sketch a path integral calculation that would justify
the expression we used earlier for the expectation value of the area of
the horizon on $\Sigma_{t_c}$:
$$
         <A> \, = \, 1/2 \times 0 + 1/2 \times A_{t_c} 
$$
where $A_{t_c}$ is the area of the horizon's intersection with
$\Sigma_{t_c}$ in the case that the collapse is triggered at $t =0$ .

Before we begin the calculation, we review the path integral approach to
ordinary non-relativistic quantum mechanics.  The basic ingredients are
paths (or ``histories'') $\gamma (t)$, i.e., curves
$\gamma:I\rightarrow\Gamma$, where $I$ is a time interval, and $\Gamma$
is the configuration space of the system in question.  The outcome of
the formalism is the assignment of {\it generalized probabilities} $P$
(positive real numbers) to certain classes $C$ of paths.  In special
situations these numbers $P$ can safely be interpreted directly as
probabilities, but in general they cannot, and recourse to a more subtle
interpretive scheme is necessary.\footnote{$^\dagger$}%
{Such interpretive schemes are described, in more or less detail, in 
[R::jim-et-al] [R::drexel] [R::isham].}
To avoid confusion with other uses of the term ``probability'', we will
refer to the number $P=\mu(C)$ associated to a particular class $C$ of
paths as the {\it quantal measure} of $C$.

By definition, a class $C$ of paths is specified by the imposition of
certain restrictions on $\gamma$.  For example, if the system has been
pre-selected to be at a certain $q_0\in\Gamma$, at time $t=t_0$, and one is
interested in where the system will be at time $t=t_1$, given that in
the intervening time period the paths accessible to the system are those
in the subset $C$, then the classes of interest may be denoted
`$C_{q_1}$' , by which we mean the class of paths that begin at $q_0$ at
$t=t_0$, belong to $C$, and terminate at $q_1$ at $t=t_1$.  The quantal
measure of $C_{q_1}$ is then (formally)
$$
  \mu(C_{q_1}) = 
  P(q_1,t_1,q_0,t_0;C)=
    N 
    | \Sigma_{\gamma\in C, \gamma (t_0)=q_0, \gamma (t_1)=q_1}
     e^{iS[\gamma]^{t_1}_{t_0}} |^2
   \eqno(E::p1)
$$
where $S[\gamma]^{t_1}_{t_0}= \int^{t_1}_{t_0} L[\gamma(t')] dt' $.  
Here, we have introduced an optional normalization constant $N$ which
can be chosen to make the measures add up to unity if desired:
$$
  \int_{q_1 \in \Gamma}   P(q_1,t_1,q_0,t_0;C) dq_1 =1      \eqno(E::p2)
$$
The intervals of definition of the curves in the class $C$ must, of
course, include any time period used in the specification of $C$.  In
the gravitational case, we will see that this can make it necessary to
consider paths that go into the future and come back to the past.

Now take another example in which all the paths under consideration obey
both $\gamma(t_0)=q_0$ and $\gamma(t_1)=q_1$ , while our interest is in
the value of $q$ at a pair of intermediate times $t_0'$ and $t_1'$
between between $t_0$ and $t_1$ ($t_0 < t_0' < t_1' < t_1$).  In such a
case, the class $C$ might be specified by a pair of characteristic
functions $C_0$ and $C_1$ of $q$ at $t=t_0'$ and $t=t_1'$,
respectively, such that $\gamma\in{C}$ iff both $C_0$ and $C_1$ take the
value $1$.  Then eq. (E::p1) becomes (we omit the optional coefficient
$N$)
$$
\eqalign{
 P =& | 
  \Sigma_{ \gamma (t_0)=q_0,  \gamma (t_1)=q_1}
  e^{iS[\gamma]^{t_0'}_{t_0}} C_0 (\gamma(t_0'))
  e^{iS[\gamma]^{t_1'}_{t_0'}}C_1 (\gamma(t_1'))
  e^{iS[\gamma]^{t_1}_{t_1'}}
    |^2
                                                       \cr   
   =&
   | \Sigma_{q_0'}\Sigma_{q_1'} \Sigma_{ \gamma (t_0)=q_0, 
      \gamma (t_1)=q_1,\gamma (t_0')=q_0', 
       \gamma (t_1')=q_1'}
 e^{iS[\gamma]^{t_0'}_{t_0}} C_0 (q_0')
 e^{iS[\gamma]^{t_1'}_{t_0'}}C_1 (q_1')
 e^{iS[\gamma]^{t_1}_{t_1'}}
 |^2
                                                     \cr
  =&
   |\Sigma_{q_0'}\Sigma_{q_1'} \Sigma_{ \gamma (t_0')=q_0', 
   \gamma (t_1')=q_1'}
   \Psi_1^* (q_1')
    e^{iS[\gamma]^{t_1'}_{t_0'}} 
   \Psi_0(q_0')
    |^2
}                                                      
$$
where we have defined the ``wave functions''
  $\Psi_0(q_0')=\Sigma_{ \gamma (t_0)=q_0,\gamma (t_0')=q_0'}
  e^{iS[\gamma]^{t_0'}_{t_0}} C_0(q_0')$ 
and
  $\Psi_1(q_1')=\Sigma_{ \gamma (t_1)=q_1,\gamma (t_1')=q_1'}
  e^{iS[\gamma]^{t_1'}_{t_1}} C_1(q_1')$
at $t=t_0'$ and $t=t_1'$ respectively.
In this way, all the conditions on the path referring to times
$t\le{t_0'}$ or $t\ge{t_1'}$ are condensed into the wave functions
$\Psi_0$ and $\Psi_1^*$.  Such wave functions will be definable whenever
there exist times $t_0'<t_1'$ such that 
the full set of conditions on $\gamma$ is the conjunction of conditions
referring solely to $t\le{t_0'}$, $t\in[t_0',t_1']$ and $t\ge{t_1'}$,
respectively. 

Both of these examples are rather special.  A more typical situation is
that in which we have an ``initial'' wave function $\Psi$ at $t=t_0$
(providing information on the behavior of $\gamma$ for times prior to
$t_0$) and our interest is in classes $C$ whose definition
refers to times $t\in(t_0,t_1)$, with no final condition on
$\gamma(t_1)$ (and therefore no relevant final wave function).
In such a case, we have
$$
  \mu(C) = \Sigma_{q_1 \in \Gamma} 
      | \Sigma_{q_0} 
        \Sigma_{\gamma \in C, \gamma(t_0)=q_0, \gamma(t_1)=q_1}
    e^{iS[\gamma]^{t_1}_{t_0}} \Psi(q_0)
    |^2 
   =
   \Sigma_{q_1 \in \Gamma}
   | \Sigma_{\gamma \in C, \gamma (t_1)=q_1}
     \A[\gamma]_{t_0}^{t_1}
   |^2
                                                      \eqno(E::p7)
$$
where we have defined  the amplitude
$\A[\gamma]_{t_0}^{t_1} \equiv e^{iS[\gamma]^{t_1}_{t_0}} \Psi(\gamma(t_0))$.

Using eq. (E::p7), one can
define an ``expectation value''
for any
physical quantity
represented by a path functional $F(\gamma)$.
(For example, for a system consisting of
a single particle moving in one dimension, $F$ might be
the position operator at
some time $t_a$, $F(\gamma)=x(\gamma(t_a))$, or the velocity 
operator at some other time $t_b$, 
$F(\gamma)= {d\over{dt}} x(\gamma(t))|_{t=t_b}$).  
Since a quantum measure does not obey the probability sum rules, the
ordinary probabilistic concept
of expectation value does not automatically carry
over, but a convenient definition for present purposes is the
following. 
Consider the range of values
that $F$ can take within $C$, which for notational simplicity
we take to be the discrete set
($f_1,f_2,f_3,\cdots$), and define the class $C_i$ by 
$$
  C_i =\lbrace \gamma \in C | F(\gamma) = f_i \rbrace    \eqno(E::ci)
$$
Obviously, $C = \cup_i C_i$. Then we can define
$$
   <F> 
    \, 
    = 
    \, 
    \sum\limits_i  f_i \ \mu(C_i) 
    \, 
    \bigg / 
    \, 
    \sum\limits_i \mu(C_i)               \eqno(E::ave)
$$
Another natural definition might be
$$
 <F> 
  \, 
  = 
  \, 
  \Re 
    \left( 
       \sum\limits_{\gamma,\tilde\gamma\in C} \A[\tilde\gamma]^* \A[\gamma] 
       F(\gamma) 
    \right) 
  \bigg /
       \sum\limits_{\gamma,\tilde\gamma\in C} \A[\tilde\gamma]^* \A[\gamma] 
  \,,   
  \eqno(E::ave-alt)
$$ 
where the sums are over all 
$\gamma,\tilde\gamma\in{C}$ such that $\gamma(t_1)=\tilde\gamma(t_1)$,
and `$\Re$' denotes ``real part of''.
In general, this differs from (E::ave), but 
using it instead of (E::ave) in the present context
would not alter our main result (E::ea2) below.

It is often
convenient [R::Raf] to consider instead of the path
$\tilde\gamma$, a new path 
$\gamma'={\tilde\gamma}^{-1}$, 
obtained from $\tilde\gamma$ by
running it from the future, $t_1$, to the past, $t_0$.
We must then remember that, if a forward running path is assigned the
amplitude $\A$, then the corresponding backward running path is assigned
the amplitude $\A^*$,
a rule that will be important to us later.
We then can write, under the conditions that led
to (E::p7),
$$
  <F> \ 
  = 
  \sum
  \limits_
    {{\scriptstyle\gamma,\gamma'\in C}\atop{\scriptstyle F(\gamma)=F(\gamma')}}
    F(\gamma)
    \A [\gamma]
    \A [\gamma'] 
  \ \bigg / 
  \sum
  \limits_
    {{\scriptstyle\gamma,\gamma'\in C}\atop{\scriptstyle F(\gamma)=F(\gamma')}}
    \A [\gamma]
    \A [\gamma']
 \eqno(E::F4)
$$
where the sums are over all $\gamma,\gamma'\in{C}$ 
with the additional
requirements that 
($i$) $\gamma(t_1)=\gamma'(t_1)$ and 
($ii$) $F(\gamma)=F(\gamma')$.
Here the path $\gamma$ is forward running from $t_0$ to $t_1$,
and the path $\gamma'$ is backward running from $t_1$ to $t_0$.
(For a backward running path $\gamma'$, we interpret ``$\gamma'\in{C}$''
and ``$F(\gamma')$'' with respect to the corresponding forward running
path, $(\gamma')^{-1}$.)

Consistency demands that all these expressions be independent of $t_1$,
as long as $t_1$ is taken to be late enough to ensure that the paths
are defined at all times that are relevant for the imposition of the
conditions $C$ and $C_i$, and this independence follows, in ordinary
quantum mechanics, directly from unitarity. [R::logan]

Now let's adapt this formalism to general relativity.
The analog of $P(q_1,t_1,q_0,t_0;C)$ in equation (E::p1) is
$$
   P(h_1,\phi_1,h_0,\phi_0;C)
   =
   \bigg|
     \sum\limits_{(g,\phi)\in C}
      e^{i S[g,\phi]^{\Sigma_1}_{\Sigma_0}}
   \bigg|^2 \,,
   \eqno(E::ps)
$$
where $h=h_{ab}$ and $\phi$ are respectively a Lorentzian metric and a
collection of non-gravita\-tion\-al matter fields on a spacetime manifold
$M$ with initial boundary $\Sigma_0=\partial_0M$ and final boundary 
$\Sigma_1=\partial_1M$, and $S$ is the action-integral 
for the gravitational 
and matter field history including the surface 
terms required to make it additive.
The sum\footnote{*}
{In theory, the sum extends over all diffeomorphism classes of manifolds
 $M$ as well as over all inequivalent ways of attaching $M$ to
 $\Sigma_0$ and $\Sigma_1$.  It seems to us unlikely, however, that any
 such functional integral cum topological sum could be well defined, and
 we expect this formal expression will give way, in the correct theory,
 to a genuine finite sum over causal sets [R::causet]
 or other discrete structures.
 In this connection, we note also that, in omitting the sum over
 topologies, we have omitted as well the complex weight-factor 
 that one can
 include with each distinct topological class.  Each consistent choice
 of weights yields a different quantum ``sector'', with the choice of
 sector determining, for example, whether a certain topological geon
 will be a boson or a fermion [R::fay-rds].}
(formal, as always)
is restricted to pairs $\gamma=(g,\phi)$ that induce the metric $h_i$ and the
matter fields $\phi_i$ on $\Sigma_i$ ($i=0,1$) and that satisfy the
additional conditions defining the class $C$ of histories [R::Hawk2].
In a cosmological setting, 
a further condition must --- plausibly --- be imposed
on $(g,\phi)$, namely a constraint on the total spacetime volume
$\int_M\sqrt{-g}d^4x$ of $M$.  This ``unimodular'' condition is
suggested by analogy with nonrelativistic quantum mechanics [R::logan]
and appears to improve the convergence of the cosmological path
integral while yielding more physically acceptable results
[R::jorma-unimod].  Even if it is adopted, however, it probably has no
influence on the asymptotically flat path integral we are concerned with
here.

For the expectation value of the history functional $F(\gamma)$, we can
use the same expressions, (E::ave) and (E::F4), as above, remembering
always that the histories must be extended sufficiently (run up to a
late enough time) so that the value of the functional $F$ can be
evaluated for each one of them.
Recall also that the sum in (E::F4) is restricted to pairs of histories
that induce the same boundary data on $\Sigma_1$ and are such that
$F(\gamma)=F(\gamma')$.
If the initial condition on $\Sigma_0$ is specified by a wave function
$\Psi_0(h_{ab},\phi)$, then the amplitude that enters into (E::F4) is, for
a forward running path $\gamma=(g,\phi)$,
$$
  \A(\gamma) = 
   e^{iS[g,\phi]^{\Sigma_1}_{\Sigma_0}} \Psi_0(h_{ab}, \phi)
  \eqno(E::amp)
$$
where $\Psi_0$ is evaluated at the configuration $(h,\phi)$ induced on
$\Sigma_0$ by the history $\gamma=(g,\phi)$.

Now let's apply the formalism to the specific case of interest: 
the evaluation of the expected area 
of the intersection of the horizon 
with the hypersurface $\Sigma_{t_c}$ 
of the previous section.
We take the initial $\Psi_0$,
defined on some initial Cauchy surface $\Sigma_0$,
to be a quasiclassical wave packet 
corresponding to the classical data describing 
the static starting configuration of metric and matter shell
described in the preceding section
(so $\Psi_0$ is centered on 
the metric $h_{ab}^0$ 
and matter fields $\phi^0$
induced on $\Sigma_0$,
by the pre-collapse classical metric and fields, 
and is approximately constant in
the neighborhood of $(h^0,\phi^0)$, corresponding to the 
static nature of the classical data).
In addition, 
of course, 
$\Psi_0$ must also describe 
the appropriate, 
non-stationary 
state  
of the quantum mechanical triggering device.

The class $C$ of histories we must consider is that of asymptotically
flat spacetimes (in order for the horizon concept to be meaningful),
and they should, in principle, extend into the future for an infinite
time (relative to infinity) for the same reason.
Once again the teleological
nature of the horizon
manifests itself,
forcing us to consider paths
that reach into the distant future, 
in order to compute the expectation value of
a functional that could naively be 
thought to depend only on information associated
with the hypersurface $\Sigma_{t_c}$.
The functional of interest to us is
$F(\gamma)=A_{t_c}$,
the area of the intersection
with the hypersurface $\Sigma_{t_c}$
of the horizon associated with the spacetime
$\gamma$.
Thus, in principle, everything is set up 
and all that remains is to use equation (E::F4) to
evaluate $<A_{t_c}>$.  

The key observation now is that
the path integral will be dominated by  
classical histories 
(i.e., ones that satisfy the classical equation of motion), 
given that we start from an almost classical initial state;
except that, 
as a consequence of the binary choice 
made by the quantum mechanical trigger,
there will be two classical histories 
(rather than only one) 
that, 
together with small fluctuations around them, 
will contribute.  
Let's call these histories $\gamma_+$ (the history with the
black hole) and $\gamma_-$ (the history without the black hole).
Then, we will have, to an excellent approximation, only four terms
contributing to the sum (E::F4).  
(More precisely we will have four {\it sets} of terms, but no harm will
result from absorbing the contributions of the fluctuations into the
amplitudes $\A(\gamma)$.)

In fact the off diagonal terms will vanish and only two terms will remain. 
Indeed, they vanish twice over, so to speak: 
first because the two histories $\gamma_+$ and $\gamma_-$ induce
macroscopically different data on the final hypersurface $\Sigma_1$,
while in (E::F4), only pairs of histories inducing equal data contribute
to the sums;
and 
second because $F(\gamma_+)\not= F(\gamma_-)$, 
while only pairs with equal $F$ contribute to the same sums.
Hence we find for the expected area,
$$
  <F>\ = 
  { \A(\gamma_+) \A(\gamma_+') \times A +
    \A(\gamma_-) \A(\gamma_-') \times 0
  \over
    \A(\gamma_+) \A(\gamma_+') +
    \A(\gamma_-) \A(\gamma_-')
  }
  =
  {1\over 2} A \,,                          \eqno(E::ea2)
$$
where $A$ denotes the area of the horizon on $\Sigma_{t_c}$ in the case
that the collapse does occur.

It is also easy to find the magnitude of the fluctuations in the area,
which we can measure by the standard deviation of $F$,
$$
  \Delta F = (<F^2> -<F>^2)^{1/2} =
  ({1\over 2}A^2-{1\over 4} A^2)^{1/2} = 
  {1 \over 2} A                                     \eqno(E::fa1)
$$
For a macroscopic black hole the fluctuations are therefore enormous.

In writing the last two equations, we have used the formulas of ordinary
probability theory.  In other settings this could be questioned, but
there is little doubt that it is appropriate here, because the only two
(sets of) histories that matter differ macroscopically from each other.
In this situation (of ``decoherence of macroscopically distinct
alternatives'') the quantum measure effectively reduces to a classical
probability measure, and one may safely utilize all of the concepts of
classical probability theory.  The vanishing of the off diagonal terms
also has the pleasant consequence that the expression (E::ave-alt) that
we introduced briefly as an alternative definition of expectation value
would have yielded exactly the same results as the definition (E::ave)
that we did use.

\section {VI. Path integral evaluation of the entropy}

Heretofore, 
we have worked under the assumption,
represented in equation (E::SBH1),
that the entropy associated to the Cauchy surface $\Sigma_{t_c}$ would
be (up to small corrections) the expectation value of the horizon area
on $\Sigma_{t_c}$.
In Section V,
we were able to evaluate this expectation value 
using an extension of the path integral formalism, 
and we found 
that it had the value one would expect, 
namely $\half\times0+\half\times{}A$, 
$A$ being the area of the horizon's intersection 
with $\Sigma_{t_c}$ 
in the case that 
the collapse occurs.  
To do this computation, 
we were forced to extend the spacetimes 
that entered into the sum over histories 
well beyond the surface 
on which we were trying to evaluate the area, 
in order to come to grips with the teleological definition 
of the event horizon.\footnote{$^\dagger$}
{A similar extension of the histories has been advocated as a way to
extend the path integral formalism to spacetimes with closed causal
curves [R::time-cycles].}

This computation of the expected area 
corroborates our earlier assumptions in part, 
but it still doesn't prove 
that the expected area can be identified with the entropy.  
In the present section 
we attempt to obtain this identification from first principles.  
That is, 
we seek an expression for entropy 
that works in general, 
but that can be shown to reduce to the average area 
in the situation at hand.  
In seeking such an expression, 
we will be led 
to a still more radical extension 
of the sum-over-histories formalism, 
namely 
the inclusion of histories that double back in time.

We will be assuming that, in more ordinary situations where the location
of the horizon is well defined (up to merely microscopic fluctuations),
the expression $-\tr\rho\log{\rho}$ yields the standard formula
$S=2\pi{}A/\kappa$ when we take $\rho$ to be the density operator 
$\rho^{ext}$
for the gravitational degrees of freedom 
of the black hole {\it exterior}
(cf. [R::chandra-symp] [R::entangle]), 
i.e. we assume that
$$
   - \tr ( \rho^{exterior}\log{\rho^{exterior}} ) 
   = 
   2\pi A / \kappa                                   \eqno(E::entangle)
$$
in such ordinary situations.  
Our aim here, 
of course, 
is not 
to justify this assumption per se, 
but only 
to show that, 
on its basis, 
we can generalize 
the area law 
to situations involving 
large scale fluctuations 
of the horizon.

Turning now to the pre-collapse situation of Sections IV and V, 
let's assume the initial state
(on some hypersurface $\Sigma=\Sigma_0$ 
corresponding to a time 
before any horizon can have formed)
to be described by a wave function $\Psi_0(X)$
taking values on the space 
of 3-metrics 
and the other field variables 
on $\Sigma_0$, i.e. $X=(h_{ab},\phi)$.
The corresponding density matrix is
$$
    \rho_0(X,Y) = \Psi_0(X) \, \Psi_0(Y)^* 
$$
At a subsequent hypersurface $\Sigma$
corresponding to a later time $t$,
the system is described by the wave function
$$
     \Psi(X) = \sum\limits_\gamma e^{iS[\gamma]} \Psi_0(\gamma(t_0))
$$
where the sum is over histories beginning in
$\Sigma_0$ and ending in $\Sigma$
and which coincide 
on $\Sigma$ 
with  $X$.
The corresponding density matrix can be expressed as
$$
  \rho (X,Y) = 
  \sum\limits_{\gamma,\gamma'}
   e^{i(S[\gamma]-S[\gamma'])}
  \Psi_0(\gamma (t_0))
  \Psi_0(\gamma'(t_0))^*
$$
$$
   =
   \sum\limits_{\gamma,\gamma'} 
   e^{i(S[\gamma]-S[\gamma'])}
   \rho_0( \gamma(t_0) , \gamma'(t_0) )                  \eqno(E::dmeval)
$$
Now in the case where we can separate 
the degrees of freedom of the system 
into two groups A and B 
associated, 
let us say, 
with two different regions of $\Sigma$, 
we can write the density matrix as
$$
   \rho ((X_A,X_B),(Y_A,Y_B))
$$
and then compute the reduced density matrix 
for the  degrees of freedom A
in the standard way:
$$
  \rho (X_A,Y_A) 
  = 
  \sum\limits_{X_B} 
  \rho((X_A,X_B),(Y_A,X_B))
$$

In the present situation, 
we will want 
to identify the variables $X_A$
with 
the region of $\Sigma$ 
{\it exterior} 
to any horizon 
that may be present 
and 
the $X_B$ 
with 
the black hole region 
{\it interior} to the horizon.  
The problem is 
that the location of the horizon  
cannot be ascertained 
from the field data $X=(h,\phi)$ on $\Sigma$.  
Rather, 
we must, 
as in the previous section, 
locate the horizon 
with respect to a spacetime 
that extends 
sufficiently 
far to the future 
of $\Sigma$.  
Here, 
we will employ the same method as there, 
but
with the added twist that, 
since 
we now need to define a density operator $\rho$
and not just a path functional $F(\gamma)$, 
we will need our paths 
to return to $\Sigma$ 
after their excursion 
into the future. 

Before continuing, 
let us pause 
to analyze a simpler, 
but analogous 
situation.
Consider 
the superposition 
of a single photon state 
with a state in which 
a pion is present together with the photon:
$$
  |\Psi> 
  =
  \sum\limits_k c_k |k> |0> + \sum\limits_{k,p} d_{k,p} |k> |p>  \,,
  \eqno(E::pipho)
$$
where the first ket in each term corresponds to 
photons and the second to pions.
The equivalent density matrix $|\Psi><\Psi|$ is
$$
  \rho =\sum\limits_{k,k'} c_k c^*_{k'} |k> |0><0| <k'| +
  \sum\limits_{k,k',p,p'} d_{k,p} d^*_{k',p'}|k> |p> <p'|<k'|
$$
$$
  + \sum\limits_{k,k',p'} c_k d^*_{k',p'}|k> |0> <p'|<k'|
  + \sum\limits_{k,k',p } d_{k,p} c^*_{k'}|k> |p> <0|<k'| \,.
$$
If we want the reduced density matrix 
for the photon,
we must trace over 
the pion degrees of freedom,
obtaining
$$
 \rho_{photon} 
 =
  \underbrace
    { \sum\limits_{k,k'} c_kc^*_{k'} |k> <k'|}_{\rho_{photon}^{0}}
  \ 
  + 
  \ 
  \underbrace
    { \sum\limits_{k,k',p}d_{k,p} d^*_{k',p}|k><k'| }_{\rho_{photon}^{1}}
  \,.
$$
Notice that $\rho_{photon}^{0}$, 
which arises from the zero-pion sector,
is still a pure state, while $\rho_{photon}^{1}$, 
which arises from the one-pion sector,
is in general highly mixed.
Notice also that 
the 
interference terms drop out upon tracing
and do not contribute 
to the reduced density matrix. 
Notice finally
that, 
if the state (E::pipho) 
corresponds, 
for example, 
to an energy eigenstate, 
then we will have 
$\rho_{photon}^{0} \rho_{photon}^{1} = 0$.
All these features will have analogs for us,
with the pion playing the role of the black hole interior, 
and the photon that of the fields outside the black hole. 

Now let's proceed to evaluate the entropy of the 
state associated with $\Sigma_{t_c}$.
First we  evaluate the density matrix at $t_c$ 
using (E::dmeval) but\footnote{*}
{The point of using such paths is not that it changes $\rho(X,Y)$
as such, which it won't, assuming unitarity.  Rather, their use lets us
locate the horizon and thereby distinguish the black hole interior from
its exterior preparatory to tracing over the degrees of freedom
of the former.  Our treatment here glosses over some
unresolved issues to which we will return at the end of this section.}
taking the
class of paths that
start at $t_0$, go to very late times,
and come back to $t=t_c$.
As before,
the result will
be dominated by classical
or nearly classical paths.
The subclass ${\cal C}$ of such paths
can, 
in our case,
be divided into two further subclasses:
the class ${\cal C}_1$  of histories 
with a black hole,
and the class ${\cal C}_2$
for which the corresponding spacetime does not contain a black hole.
Then we have
$$
  \rho_{t_c} (X,Y) 
  \approx  
  \sum\limits_{\gamma,\gamma'\in {\cal C}} 
    e^{i(S[\gamma]-S[\gamma'])}
    \rho_0( \gamma(t_0) , \gamma'(t_0) )
$$
$$ 
  =
  \left(
    \sum\limits_{\gamma,\gamma'\in {\cal C}_1} 
    +
    \sum\limits_{\gamma,\gamma'\in {\cal C}_2} 
    +
    \sum\limits_{\gamma\in {\cal C}_1,\gamma'\in {\cal C}_2} 
    +
    \sum\limits_{\gamma\in {\cal C}_2,\gamma'\in {\cal C}_1} 
  \right)
    e^{i(S[\gamma]-S[\gamma'])}
     \rho_0( \gamma(t_0) , \gamma'(t_0))
  \eqno(E::dmeval2)
$$

Now we claim that tracing over the degrees of freedom 
associated with 
the black hole interior yields 
(in analogy with our photon-pion example) 
a reduced density matrix which
contains no interference terms.
That is, 
the instruction 
to take the trace 
over the degrees of freedom
residing within the black hole
will force us, 
in the case of the last two sums in (E::dmeval2), 
to compute the inner product 
between two 
partial wave functions, 
one of which corresponds 
to a flat metric in a macroscopic, spherical region 
(the region inside the black hole in the case where the collapse occurs)
and the other of which corresponds 
to the empty set
(the region inside the black hole in the case where no collapse occurs).
This inner product should certainly vanish, 
and therefore 
there will be no contributions 
from the last two sums in (E::dmeval2).
Thus we get for the reduced density matrix 
for the black hole exterior:
$$
 \rho_{t_c}^{ext} (X,Y) 
 = 
 p \, \rho_{t_c}^{(1)} (X,Y) 
 +
 q \, \rho_{t_c}^{(2)} (X,Y)                  \,,\eqno(E::rho12)
$$
where the superscripts 1 and 2 indicate the contribution 
arising from the black hole and the no black hole sectors respectively 
and where $p$ and $q$ indicate the corresponding probabilities 
(which we took in Section V to be $p=q=1/2$).
What is equally important, we claim that 
$\rho^{(1)}_{\tc}$ and  $\rho^{(2)}_{\tc}$ 
are orthogonal:
$$
    \rho^{(1)}_{\tc} \rho^{(2)}_{\tc} 
  = \rho^{(2)}_{\tc} \rho^{(1)}_{\tc} 
  = 0                                      \eqno(E::orthog)
$$
The reason, as before, is that the two correspond to entirely different
types of 
(exterior)
geometries, 
the former contains a flat region with a big spherical hole, 
the latter contains the same flat region with the hole filled in. 
 
Now let's evaluate the entropy (E::S1)
associated with $\rho_{t_c}^{ext}$. 
To that end 
we express (E::S1) as a series 
in terms of 
the traces of powers of $\rho$.  
Letting $\sigma:=1-\rho$, we have in general 
$$
   - \rho \log{\rho}
   = 
   - \rho \log(1 - \sigma)
   =
   \sum\limits_{n=1}^\infty { \rho \sigma^n \over n}
   =   
   \sum\limits_{n=1}^\infty { \rho (1 - \rho)^n \over n }
   =
   \sum\limits_{n=1}^\infty 
     \sum\limits_{k=0}^n    
       {n \choose k}
       {(-1)^k \over n} \rho^{k+1} \,.
$$
Hence\footnote{*}%
{We can't amalgamate the sums by collecting all the multiples of
$\tr\rho^n$, for each $n$, into a single term, because the series does
not converge absolutely.}
$$
   \tr(-\rho\log{\rho})
   = 
   \sum\limits_{n=1}^\infty 
     \sum\limits_{k=0}^n    
       {n \choose k}
       {(-1)^k \over n} \tr(\rho^{k+1}) \,.       \eqno(E::series)
$$
But from (E::orthog) we have
$$
  \tr(\rho^{ext})^n
  = 
  \tr ( p \rho^{(1)} + q \rho^{(2)} )^n
  = 
  \tr (p \rho^{(1)})^n + \tr (q \rho^{(2)})^n
$$
whence
substituting $\rho^{ext}_{t_c}$ into (E::series) yields
for the entropy
$$
\eqalign{
  S_{t_c}
  &=
    \tr(- \rho^{ext} \log{\rho^{ext}})
    \cr
  &=
    \tr(- p \rho^{(1)} \log{p\rho^{(1)}})  
    +
    \tr(- q \rho^{(2)} \log{p\rho^{(2)}})  
  \cr
  &=
    p \tr ( - \rho^{(1)} \log \rho^{(1)} ) 
    +
    q \tr ( - \rho^{(2)} \log \rho^{(2)} )
    +
    p\log(p^{-1}) + q\log(q^{-1})
  }
$$
That is to say, 
we have found that 
the entropy on $\Sigma_{t_c}$ is --- 
up to the correction term 
$S'=p\log(p^{-1})+q\log(q^{-1})$,
which is the entropy associated with 
the unresolved alternative 
of having or not having a black hole
on $\Sigma_{t_c}$  ---
the appropriately weighted average of the entropies 
of the two cases,
``horizon present'' and 
``horizon absent''.
In view of (E::entangle), 
we thus obtain the desired result
$$
   S_{t_c} = p \times {2\pi\over\kappa} A + q \times 0 + S' \,,  \eqno(E::ans)
$$
(where, in our specific example in Sections IV and V, we had $p=q=1/2$).

Before concluding this section, we wish to call attention to certain
shortcomings of our definition of a reduced density matrix $\rho^{ext}$,
related to the type of zigzag path we were led to employ.  The matrix
$\rho(X,Y)$ 
has two slots corresponding to 
the two paths $\gamma$ and $\gamma'$ 
in (E::dmeval).  
If we take both  $\gamma$ and  $\gamma'$
to go to the future and return to $\Sigma_{\tc}$ then both $X$ and $Y$
can be endowed with information on the horizon location 
(they will become in effect 
not pairs $X=(h,\phi)$, 
but {\it triples} $X=(h,\phi,b)$, 
where $b$ is a ``marking'' 
telling us 
where the black hole region sits within $\Sigma$).  
In this extended sense (which is the
one we had in mind in writing most of this section), the $\rho^{ext}(X,Y)$
that results from tracing out the interior degrees of freedom will be
hermitian and positive, but not necessarily normalized to unit trace.
On the other hand, starting from the alternative definition (E::ave-alt)
of $<F(\gamma)>$, 
we can form a slightly different $\rho^{ext}$,
of which
only the second slot partakes of a zigzag path.
This $\rho^{ext}$ contains all the information needed to determine 
$<F(\gamma)>$ for any functional of the {\it exterior} fields in the
neighborhood of $\Sigma_{\tc}$.  It is normalized, but not necessarily
hermitian, and it yields,
in our quasiclassical approximation,
exactly the same value (E::ans) 
for $\tr(-\rho\log\rho)$.  
In addition
further small variations of the definitions are possible.
Which of
these $\rho$'s, if any, furnishes the appropriate description of
conditions outside the black hole?  And does 
at least
one of them evolve
autonomously, as needed for the general proof of entropy increase
offered in [R::Raf2]?  
Clearly, the answers to these questions
lie in the full quantum theory of gravity.
Indeed, one can hope that difficulties of principle such as these
(together with some of the options for resolving them) will provide
important clues to the characteristics that the fundamental theory of
quantum gravity will have to possess.

The developments in this section have been based on histories that
begin at $\Sigma_0$, proceed into the distant future, and then double
back in time, returning to $\Sigma_{\tc}$.  In connection with their
use, it may be interesting to ask what is the nature of the wave
function that one obtains by evolving $\Psi_0$ via such histories, or
more specifically by evolving via subsets for which the collapse either
is or is not triggered.  The initial state $\Psi_0$ describes a static
geometry with nothing going on but the metaphorical ticking of the clock
of the quantum trigger.  Evolved forward to $\Sigma_{\tc}$, this wave
function is still the same, except that the trigger is closer to its
moment of decision.  What will we get 
at $\Sigma_{\tc}$
by, for example, continuing the
evolution forward via the histories 
undergoing
collapse, and then evolving
backward to $\Sigma_{\tc}$?  One can show (assuming unitarity)
that the result will be (up to small corrections) the
same, except that the trigger will be in the
state that would guarantee collapse 
(for example the correlated EPRB photons, if they are already in flight,
will have positive helicity), 
and
vice versa for the histories with no collapse.  To return to such a
pre-collapse
state from a late time 
black hole 
is of course
highly ``anti-thermodynamic'' behavior --- an example of the
``Umkehreinwand'' with real practical significance.  If we continue to
evolve backward, we get states of increasingly more
bizarre anti-thermodynamic content.  For example, if the decision
mechanism utilized a photon impinging on a half silvered mirror, then we
would get a superposition of the photon emerging from its true source,
with a photon emerging backward from whatever it is that absorbs the
photon in the case that the collapse is to be triggered.

\section{VII. Further thoughts}

We have described 
two ``Gedankenexperimenten'' 
in which 
macroscopically large fluctuations of the entropy 
are induced by 
the amplification of 
microscopic quantum events.  
Along with the possibility 
of large entropy increases 
in these experiments 
comes the possibility 
of large entropy {\it decreases},
which, 
however, 
are still consistent with the second law of thermodynamics 
in the sense that, 
on average, 
they are balanced by the increases.  

Now, 
these large entropy decreases 
are not 
the type of
``reversal of the thermodynamic arrow of time''
that would take place, 
if,
for example, 
all the air molecules in a room 
suddenly decided 
to migrate into one corner.  
Rather, 
they occur when a
certain kind of 
``quantal superposition'' 
of a low entropy alternative
with a high entropy alternative
resolves itself into 
one or the other possibility 
in a manner reminiscent of ``state vector reduction''.

The question arises, 
Is the entropy $S$ 
that fluctuates in these examples 
a ``Gibbs entropy'' or a ``Boltzmann entropy'', 
does it refer to an ``ensemble'' or to an individual system?  
If we were think of $S$ as 
the Gibbs entropy of an ensemble 
then 
its downward fluctuations could seem 
subjective and unreal 
(and in fact would not occur at all if we kept the ensemble intact).  
On such a view, 
the entropy 
at time $\tc$
in the black hole example 
would be, 
instead of (E::ans),
either $2\pi A/\kappa$ or $0$ 
depending on whether the horizon was
``really present'' or not.  
In neither case
would it fluctuate,
since each element of the ensemble
would represent an essentially classical spacetime 
exhibiting the normal growth of horizon area with time.
The trouble with this point of view is 
that there appears to be no physical basis on which to distinguish 
the one alternative from the other --- at least on the basis of anything
existing at time $\tc$.  Rather the entropy seems to derive from 
the system's {\it potential} to evolve in two very different ways.
Thus, it seems to us better 
to view (E::ans) as 
a kind of Boltzmann entropy, 
but with the peculiarity that 
it is not in any evident sense 
a measure of the number of microscopic complexions 
of the corresponding macrostate.
(At best it is an average of such complexion measures.)  
On this view, the entropy fluctuations are objective and ``real'', 
even if 
the traditional formula $S=\log{N}$ 
fails to tell the whole story.

In any case,
both the above views 
share the feature that 
the entropy at time $\tc$ 
can be quite large 
even though the spacetime is flat. 

Indeed, 
we believe that
the reflexions presented in the present work
have led to a coherent, 
internally consistent
picture, 
that imputes
a well-defined and computable entropy
to hypersurfaces like $\Sigma_{\tc}$.
This picture bears out
the conception that, 
if an event horizon is present with some probability,
then a corresponding entropy must also be present. 

We believe that the existence of this kind of entropy poses a severe
challenge to anyone attempting to base a theory of quantum gravity on the
method of ``canonical quantization''.
Let's suppose for a moment that we have the correct
theory of quantum gravity, and moreover
that it is formulated in a canonical manner;
that is,
the theory
singles out some choice of classical
canonical variables and identifies them
to corresponding objects in the quantum theory.
(These variables could be 
the induced metric and extrinsic curvature of the ADM formalism, 
the Ashtekar variables, 
or the string degrees of freedom
to name some of the 
possibilities 
that have been considered in a canonical context.)
If, moreover, 
we have also the correct theory for 
non-gravitational matter, 
either in a form  
truly unified with gravity 
as in String Theory,
or not, 
then we 
should 
be able to evaluate any desired quantity for a fully 
specified state of the system.
And for any 
``macroscopically fully specified'' 
state of the system 
(that is, one for which the macroscopic degrees of freedom 
are fully specified), 
we should, 
in accordance with the principle 
of
the 
statistical mechanical origin of thermodynamics,
be able to evaluate all the
relevant thermodynamical quantities.

Consider now the problem of computing, 
from first principles, 
the entropy associated with 
a given configuration 
of gravitational and ``matter'' variables.  
This should be possible to do 
(through the introduction of suitable ensembles 
constructed by coarse graining starting from the given state), 
as long as 
at least 
the macroscopic degrees of freedom
are fully specified. 
Moreover the answer should 
agree
with the 
known thermodynamical result 
that
assigns a contribution 
to the entropy 
equal to $(2\pi/\kappa)$ of the horizon area 
to those situations involving a black hole. 
Now consider
the 
state
associated with the hypersurface
$\Sigma_{t_c}$ in section III. 
The classical metric
and
canonical momentum 
on $\Sigma_{t_c}$
are completely specified,
and so are the  
macroscopic degrees of freedom 
of the matter.
(Note, moreover, that even though 
this could not be said to be a strictly stationary situation, 
the only thing that is changing with time 
is 
the internal clock
associated with the quantum mechanical device.
Everything else
is perfectly static,
and in particular,
the macroscopic metric around $\Sigma_{t_c}$ 
possesses a timelike Killing Field.)
Thus, the theory should assign 
to this macroscopic configuration 
a pure or mixed state
of the 
underlying 
quantum mechanical variables,
and therefore
the entropy
should be fully determined. 
However, 
as we have seen,
the situation on $\Sigma_{t_c}$ is such 
that 
the entropy
lies in between two very different values,
depending on a small detail
concerning  an energetically insignificant 
degree of freedom of the matter fields 
(related to the quantum mechanical phases in the triggering device). 
It follows that 
{\it the purported  theory of quantum gravity 
must be extremely sensitive 
to such things as 
the phases of the matter degrees of freedom} 
if it is to successfully reproduce
the result obtained in section IV. 
It is difficult to envision
a canonical theory
developing in such a way as to incorporate 
this very strange feature which seems necessary 
if 
the correct result 
is to be obtained 
from it.

The following objections can be 
(and have been) 
advanced 
to counter our arguments:
(i) Entropy should be defined only for stationary (equilibrium) states. 
(ii) The black hole entropy could be associated with the
area of the apparent horizon rather than the event horizon,
and 
the former {\it is} determined from data on $\Sigma$. 
(iii) It is unreasonable to expect the theory of 
quantum gravity to  solve also 
the ``quantum measurement problem'',
which evidently is a central feature of the 
situation described in section III.
We feel all of these criticisms are unwarranted, 
as we now explain.

(i) This could be argued for the thermodynamical entropy but not for 
the statistical mechanical entropy which is what a fundamental 
theory should yield 
(through the introduction of suitable coarse grainings, etc.).
Moreover
the situation 
on $\Sigma_{t_c}$
is 
for all practical purposes 
static,
and 
it seems 
hard to deny that 
in such a case
we should have 
a well defined
thermodynamical entropy. 
%
%
Finally, by
arguing that there are situations in which entropy is not defined,
one would be calling into question the status of the second law 
as an
argument against perpetual motion machines.
In relation to black holes in particular, one loses the thermodynamic
interpretation of the classical area increase theorem if one limits the
application of the entropy concept to stationary black holes (and
similarly for that portion of the ``first law'' that covers
perturbations to nearby {\it non-stationary} solutions).

(ii) 
First, our best evidence for the thermodynamic properties of black
holes is based on the event horizon, and not the apparent horizon.  Of
course, the two coincide for stationary  black holes, but in the
non-stationary case, we have for the event horizon, the classical area
increase theorem as well as that portion of the so called first law that
states that variations in the energy are proportional to variations in
the area, even for non-stationary perturbations [R::TdA].  In contrast,
the apparent horizon jumps discontinuously in dynamical spacetimes,
which it is hard to believe the physical entropy would do;  
and, to our
knowledge, the possibility of sudden {\it decreases} in its area has not
been ruled out.  Second, the intuitive connection between entropy and
hidden information is lost for the apparent horizon, given that one
doesn't even know in general whether it divides spacetime into interior
and exterior regions.  (In fact, due to the discontinuous jumps, the
apparent horizon will {\it not} divide every hypersurface of a general
foliation.)  Third, the apparent horizon seems too tightly tied to a
smooth spacetime metric for the concept to survive in a fundamental
theory of quantum gravity.  Already in simple quantum models
[R::parker], 
it loses its property of being (locally) achronal
(which is crucial for the
autonomous evolution of the exterior region) and one may well doubt
whether ``off shell'' it could escape being scattered in bits and pieces
all over spacetime.  In contrast, the notion of a black hole as the
region causally cut off from infinity generalizes even to discrete
settings, and by definition, the region outside is causally independent
of the region inside, at least kinematically. 
Finally, the apparent horizon itself is not local in time (unless with
respect to some distinguished foliation of the spacetime manifold), and
in this respect is no better off than the event horizon.  
Indeed,
the Schwarzschild black hole 
spacetime
has been shown to possess Cauchy hypersurfaces that 
contain no trapped surfaces [R::Vivek]
and therefore contain no apparent horizon
that could be recognized from canonical initial data.\footnote{$^\dagger$}
{We don't know whether adding information from the past (but not the
 future) would suffice.  If not, then the apparent horizon would
 actually be worse off than the event horizon, for which, at least,
 knowledge of the {\it future} always suffices!}
This
looks like
an unsurmountable obstacle to 
the idea that the black hole entropy should be 
associated with the
area of the apparent horizon
in preference to the event horizon.

(iii) The significance of the proposed example is
precisely that it strongly 
advances the opposite
point of view
--- at least if the words ``measurement problem'' are construed in a
sufficiently general sense.
The fact that issues related to 
the
interpretation of quantum mechanics 
creep
into the evaluation of 
a physical quantity associated with times earlier 
than those at which the 
pertinent
``measurement'' is to occur
(something that to our knowledge does not occur in
nongravitational physics; cf. however [R::osgood]) 
is, in our opinion, a further indication that 
some change in the formal structure of quantum mechanics 
will be a necessary step on the journey to
a satisfactory 
theory of quantum gravity.\footnote{*}
{Such views have long been advocated by R.~Penrose, among others.  In
 connection with the notion expressed in [R::pen] that gravity might
 ``collapse the wave function'', we remark that in our example, the
 difference between the quantum state corresponding to collapse and that
 corresponding to non-collapse is gravitationally negligible at the time
 $t_{\tc}$.  Certainly the ``difference in spacetime curvature'' is no
 greater than that found in everyday quantum interference experiments.}
We expect in particular, that one consequence of this change will be
the emergence of an interpretation of the existing formalism not tied to
the idea of measurement on a hypersurface (or any other pre-specified
spacetime region).

The preceding arguments are of course very far
from a proof that 
canonical theories are
wrong,
but they seem 
at least 
to pose
a severe challenge to any 
such
theory that is
presented as the correct theory of quantum
gravity and which in particular is claimed to successfully 
evaluate the entropy of a black hole.
In this respect, 
it is worth noting
the relative ease with which 
(modulo the subtleties discussed above) 
the path integral
yields
the intuitively correct result 
for
the situation in question, 
of course 
under the assumption that 
it yields the standard result $(2\pi/\kappa)A$ 
in standard situations.

This research was partly supported by NSF grant PHY-9600620, by a
grant from the Office of Research and Computing of Syracuse University
and by Proyect IN 121298 DGAPA-UNAM.

\ReferencesBegin


[R::forks]
R.D.~Sorkin,
``Forks in the Road, on the Way to Quantum Gravity'', talk 
   given at the conference entitled ``Directions in General Relativity'',
   held at College Park, Maryland, May, 1993,
   {\it Int. J. Th. Phys.} {\bf 36}: 2759--2781 (1997)   
   \eprint{gr-qc/9706002}

[R::stoch-manifold]
 R.D.~Sorkin, 
``Stochastic Evolution on a Manifold of States'',
  {\it Ann. Phys. (New York)} {\bf 168}: 119-147 (1986)

[R::meso-def]
Marco Roncadelli, 
 ``Random Path Formulation of Nonrelativistic Quantum Mechanics'', 
  in
  {\it Lectures on Path Integration: Trieste 1991},

  eds. H.A.~Cerdeira et al.,
  pp. 517-539
  (World Scientific 1993)

[R::ref-on-increase] 
Woo Ching-Hung,
``Linear Stochastic Motions of Physical Systems'',
 Berkeley University Preprint, UCRL-10431 (1962)

G\"oran Lindblad,
``Completely Positive Maps and Entropy Inequalities''
  {\it Commun. Math. Phys.} {\bf 40}: 147-151 (1975)

[R::Raf2]
 R.D.~Sorkin, 
``Toward an Explanation of Entropy Increase 
  in the Presence of Quantum Black Holes'',
  {\it Phys. Rev. Lett.} {\bf 56}: 1885-1888 (1986)

[R::OPen]
O.~Penrose, {\it Foundations of Statistical Mechanics}
(Pergamon Press, Oxford, 1970).

[R::Isr]
For the classical analysis of collapsing shells see: 
W. Israel, Nuovo Cimento {\bf 44 b}, 1 (1966);
W. Israel, Phys. Rev. {\bf 153}, 1388 (1967).

[R::chandra-adelaide] 
 R.D.~Sorkin,
``The Statistical Mechanics of Black Hole Thermodynamics'',
  in R.M. Wald (ed.) {\it Black Holes and Relativistic Stars}, 
  (U. of Chicago Press, 1998)
  \eprint{gr-qc/9705006}.

R.D.~Sorkin,
``How Wrinkled is the Surface of a Black Hole?'',
  in David Wiltshire (ed.), 
  {\it Proceedings of the First Australasian Conference on General
       Relativity and Gravitation}, 
  held February 1996, Adelaide, Australia, pp. 163-174
  (University of Adelaide, 1996)
  \eprint{gr-qc/9701056}.


[R::jim-et-al] 
J.B.~Hartle, ``Spacetime Quantum Mechanics and the Quantum 
 Mechanics of Spacetime'',
 in B.~Julia and J.~Zinn-Justin (eds.),
 {\it Les Houches, session LVII, 1992, Gravitation and Quantizations}
 (Elsevier Science B.V. 1995),
 and references therein.

[R::drexel]
  R.D.~Sorkin,
``Quantum Measure Theory and its Interpretation'', in
    D.H.~Feng and B-L~Hu (eds.), 
    {\it Quantum Classical Correspondence:  Proceedings of the $4^{\rm th}$ 
     Drexel Symposium on Quantum Nonintegrability},
       held Philadelphia, September 8-11, 1994, pages 229--251
    (International Press, 1997)
    \eprint{gr-qc/9507057}.

[R::isham] 
C.J.~Isham, ``Quantum Logic and the Histories Approach to Quantum Theory'',
  {\it J. Math. Phys.} {\bf 35}: 2157-2185 (1994)
  \eprint{gr-qc/9308006}.

[R::Raf]
See for example  
Sukanya Sinha and Rafael~D.~Sorkin,
``A Sum-over-histories Account of an EPR(B) Experiment'',
   {\it Found. of Phys. Lett.} {\bf 4}:303-335 (1991).

[R::logan]
 R.D.~Sorkin, 
 ``A Modified Sum-Over-Histories for Gravity'',
   reported in
  {\it 
   Proceedings of the International Conference on Gravitation and Cosmology, 
   Goa, India, 14-19 December, 1987},
   edited by 
   B.~R. Iyer, Ajit Kembhavi, Jayant~V. Narlikar, and C.~V. Vishveshwara,
   see pages 184-186 in the article by 
   D.~Brill and L.~Smolin: 
   ``Workshop on quantum gravity and new directions'', pp 183-191 
   (Cambridge University Press, Cambridge, 1988);

 R.D.~Sorkin, 
``On the Role of Time in the Sum-over-histories Framework for Gravity'',
    paper presented to the conference on 
    The History of Modern Gauge Theories, 
    held Logan, Utah, July 1987, 
    published in  
      {\it Int. J. Theor. Phys.} {\bf 33}:523-534 (1994).

[R::causet]
L.~Bombelli, J.~Lee, D.~Meyer and R.D.~Sorkin, 
``Spacetime as a Causal Set'', 
  {\it Phys. Rev. Lett.} {\bf 59}:521-524 (1987).

[R::fay-rds]
 H.F.~Dowker and R.D.~Sorkin,
``A Spin-Statistics Theorem for Certain Topological Geons'',
  {\it Class. Quant. Gravity}{\bf 15}: 1153-1167 (1998)
  \eprint{gr-qc/9609064}.

[R::Hawk2]
S.W.~Hawking in 
 {\it General Relativity, an Einstein Centenary Survey}, 
 eds. S.W. Hawking and W. Israel, 
(Cambridge U. Press, London, 1979).

[R::jorma-unimod]
 Alan Daughton, Jorma Louko and Rafael D.~Sorkin,
``Instantons and unitarity in quantum cosmology with fixed four-volume'',
 {\it Phys.~Rev. D}{\bf 58}: 084008 (1998) 
 \eprint{gr-qc/9805101}.

[R::time-cycles]
J.L.~Friedman, N.J.~Papastamatiou and J.Z.~Simon, ``Failure of unitarity
  for interacting fields on spacetimes with closed timelike curves,'' 

  {\it Phys. Rev. D} {\bf 46}:4456 (1992)

[R::entangle]
 R.D.~Sorkin, 
``On the Entropy of the Vacuum Outside a Horizon'',
  in B. Bertotti, F. de Felice and A. Pascolini (eds.),
  {\it Tenth International Conference on General Relativity and Gravitation
  (held Padova, 4-9 July, 1983), Contributed Papers}, 
  vol. II, pp. 734-736
  (Roma, Consiglio Nazionale Delle Ricerche, 1983)
  %

[R::TdA]
R.D.~Sorkin and M. Varadarajan,
``Energy Extremality in the Presence of a Black Hole'',
  {\it Class. Quant. Gravity} {\bf 13}: 1949-1970 (1996) 
  \eprint{gr-qc/9510031}

D.~Sudarsky and R.~Wald, 
``Extrema of Mass, Stationarity, and Staticity, and Solutions to the
    Einstein-Yang-Mills Equations'',
 {\it Phys.~Rev.~D} {\bf 46}: 1453 (1992)

[R::parker]
Leonard Parker, ``Semi-infinite throat as the end-state of two-dimensional
   black hole evaporation'',
   {\it Phys. Rev. D} {\bf 52}: 3512-3517 (1995)

[R::Vivek]
R. M. Wald and V. Iyer, Phys. Rev. D, R3719, (1991).

[R::osgood]
 R.D.~Sorkin, 
``Problems with Causality in the Sum-over-histories Framework
      for Quantum Mechanics'',
   in A. Ashtekar and J. Stachel (eds.), 
  {\it Conceptual Problems of Quantum Gravity} 
   (Proceedings of the conference of the same name, 
       held Osgood Hill, Mass., May 1988), 217--227 
   (Boston, Birkh\"auser, 1991)

[R::pen]
 Penrose R. {\it The Emperor's New Mind: 
 Concerning Computers, Minds, and the Laws of Physics}, 
 (Oxford Univ. Press, Oxford, 1989).

[R::exp]
See for example R. M. Wald, 
{\it Quantum Field 
Theory in Curved Spacetime 
and Black Hole Thermodynamics}
(University of Chicago Press, 1996), and references therein.

[R::Callan]
 See for example 
 H. Callan, {\it Thermodynamics and an Introduction to Thermostatistics}
 (John Wiley \& Sons, Ney York, 1985).

[R::Shan]
 C.E.~Shannon and W.~Weaver, 
 {\it The Mathematical Theory of Communication}
 (Univ. of Illinois Press, Urbana, 1949).

\end


(prog1    'now-outlining
  (Outline 
      "
     "......"
      "
   "\\\\message" 
   "\\\\section" 
   "\\\\appendi"
   "\\\\Referen"	
   "\\\\Abstrac" 	
      "
   "\\\\subsectio"
   ))